\documentclass[prl]{revtex4}%

\usepackage{epsfig}
\usepackage{graphics}

\begin{document}

\title{Quantum Gravitational Uncertainty of  Transverse Position}

\author{Craig J. Hogan}
\altaffiliation{also  Max-Planck-Institut f\"ur Astrophysik, Karl-Schwarzschild-Str. 1, D-85748 Garching bei M\"unchen, Germany}
\affiliation{University of Washington, Departments of Physics and Astronomy, Seattle WA 98195}

\begin{abstract}
It is argued that holographic  bounds on the information content of spacetime might be directly measurable.
A new uncertainty principle is conjectured to arise from   quantum indeterminacy of nearly flat spacetime: {\em  Angular orientations of null trajectories of spatial length  $L$ are uncertain, with standard deviation in each transverse direction} $\Delta \theta> \sqrt{l_P/L}$, where  $l_p$ denotes the Planck length.   It is shown that this angular uncertainty  corresponds to the information loss and nonlocality that occur if 3+1-D spacetime has a holographic dual description in terms of  Planck-scale waves on a 2+1D screen with encoding close to the Planck diffraction limit, and    agrees with covariant holographic entropy bounds on total number of degrees of freedom. The spectrum and spatial structure of  predicted quantum-gravitational ``holographic noise'' are  estimated to be directly measurable over a broad range of frequencies using   interferometers with current technology. 
\end{abstract}
\maketitle
\section{Introduction}
Evidence has accumulated for some time that spacetime is holographic:  the number of fundamental degrees of freedom of a system in spacetime, including the quantized degrees of freedom of the spacetime itself, scales with area rather than volume, as would be expected in quantum field theory.  The laws of black hole thermodynamics\cite{Bekenstein:1972tm,Bardeen:gs, Bekenstein:1973ur,Bekenstein:1974ax}  revealed that  the area of the event horizon of a black hole behaves like thermodynamic entropy.  With the discovery of Hawking radiation\cite{Hawking:1975sw} the identification of geometric entropy with quantum degrees of freedom of evaporated particles became explicit\cite{'tHooft:1985re}. Later, actual black hole quantum states  were enumerated in detail (in the low temperature limit) using explicit constructions of extremal  black holes in higher dimensions\cite{Strominger:1996sh}. In the case of black   holes the total  entropy is known with surprising precision:  the number of bits of information needed to completely specify the state of a black hole, including all the configurations of fields and spacetime quanta, is given by $S= A/l_P^24\ln 2 $, where $A$ is the area of the event horizon, $l_P=\sqrt{\hbar G_N/c^3}= 1.616\times 10^{-33} {\rm cm}  $ is the Planck length, and $G_N$ is Newton's gravitational constant. (In the following, we set $\hbar=c=1$ unless otherwise noted.)

Some understanding of these effects, and indications that a consistent description exists of spacetime as a unitary, holographic quantum system, came about from the realizations that the field equations of classical general relativity can be derived from a boundary theory rather than extremizing the volume integral of Einstein-Hilbert action\cite{Jacobson:1995ab}, and that the externally observable entropy of a volume of quantized fields, when traced over unobservable internal degrees of freedom, scales as area
\cite{Bombelli:1986rw,Srednicki:1993im,Yarom:2004vp}.
A holographic principle has been conjectured to apply not just to black holes, but to any spacetime \cite{'tHooft:1993gx,Susskind:1994vu,'tHooft:1999bw,Bigatti:1999dp}. 
Covariant  holographic  entropy bounds  generalize   to other spacetimes, apparently reflecting  a deep universal feature of quantum gravity that associates degrees of freedom with null surfaces\cite{Bousso:2002ju,Padmanabhan:2007en}. These properties may be clues to the construction of an effective quantum theory out of which classical spacetime behavior emerges as a large-scale behavior\cite{Padmanabhan:2007tm}.
Fully holographic theories have now been demonstrated, in which 
a system of quantum  fields  and dynamical gravity in $N$ dimensions is dual  to  a system of quantum fields in $N-1$ classical 
 dimensions\cite{Maldacena:1997re,Witten:1998zw,Aharony:1999ti,Alishahiha:2005dj,Horowitz:2006ct}. 
  
Direct  experimental tests of these ideas have however seemed out of reach, partly because the Planck scale is so incredibly tiny   (an exception being models invoking relatively large extra dimensions on submillimeter scales, that  bring the quantum gravity effects closer to experimental capabilities\cite{Adelberger:2003zx}), but also because
 there has been no holographic description of nearly-flat 3D space that explicitly displays its quantum degrees of freedom in order to allow predictions of observable phenomena that can be tested in experiments.  The holographic bounds give a maximum Hilbert space dimension for 3D quantum gravity,  but do not tell us what the observable eigenstates or degrees of freedom  of the system look like ``from inside''. Measuring such quantum-gravitational phenomena would help bridge the conceptual gap between string-based quantum gravity and the limit of classical spacetime. 

This paper   formulates a concrete conjecture about the character of 3D quantum spacetime eigenstates that permits   predictions of observables. Basic holographic optics are invoked to motivate a simple ``holographic uncertainty principle'' associated with the orientation of null trajectories in   3D space.     The  conjectured holographic uncertainty principle corresponds to an irreducible   uncertainty of observable states of   spacetime itself, and therefore of observable nonlocal relative spatial positions  in all  systems, in directions transverse to macroscopic spatial separation vectors.    The uncertainty corresponds to a distance-dependent limit on the  number of distinguishable directions in 3-space  that leads to the holographic scaling of entropy with area rather than volume: whole systems  then have the same number of degrees of freedom as allowed by holographic entropy bounds, but remain consistent with high resolution local physics near any observer.   If this principle indeed captures the holographic behavior of 3D quantum spacetime states, quantitative estimates suggest that  it may be possible  to study   holographic  degrees of freedom  of spacetime directly using technology adapted from existing gravitational wave experiments. 
Laboratory-scale interferometers may be capable of detailed studies of a new kind of quantum-gravitational ``noise'' arising from the limited information content of spacetime.  Predictions for the character of the noise and its spectrum are described  in more detail in  \cite{Hogan:2007ci,Hogan:2007hc}. 

\section{Description of   space as  a Planck-wave hologram}
Suppose that the 3D world has a description as a hologram on a 2D surface.  The 3D positional information is encoded as an interference pattern on a surface at Planck scale  resolution. Imagine for the moment that the 2D hologram is illuminated with a reference beam of Planck wavelength ($l_P$) radiation,  and a virtual 3D world appears.  The  3D world becomes increasingly blurry as the distance from the screen increases. This happens in such a way that the information encoded on the 2D screen viewed through a pupil of diameter $D$, about $(D/l_P)^2$ bits, matches the number of observably independent pixels in the 3D world, so the two descriptions have the same amount of information.  At a large distance $L$ from the screen, the diffraction limited spot size is $\Delta x_D\simeq Ll_P/D$, so   a surface subtending a large solid angle in the 3D world  has $(L/\Delta x_D)\simeq (D/l_P)$ distinguishable angular  pixels in each transverse direction, independent of $L$.  The screen is able to encode $\simeq (D/l_P)^2$ different position pixels in the 3D world, ranging from Planck scale 3D pixels near the screen to much coarser ones far away. Near the screen there is little directional information but good transverse spatial resolution; far away, the situation is reversed.

Now imagine that the 3D world is not virtual at all, but our real world. There must be blurring of distant spatial structure if there really is a 2D holographic description, otherwise the 3D world would have too much data; but now we interpret the 2D holographic wave as a quantum-mechanical wavefunction and the 3D world that emerges from it as an approximation to nearly-flat classical spacetime. 
The 3D ``insider's view''  of the hologram represents an observable classical spacetime, but it is blurry in the sense of having a quantum indeterminacy of angle and transverse position. 

We anticipate that the fundamental objects are actually null trajectories and light sheets, so the blurring is entirely in angle or transverse position, and is not detected in radial distances between events with null separation.
   The blurring is both nonlocal and observer-dependent, in the sense that the  spacetime eigenstates (the character of which determines   blurring) depend on the location and motion of  the frame in which they are measured. 
 Although Planck-fidelity observations can be made anywhere, it should not be possible to make arbitrarily precise relative position  measurements everywhere in space on a single branch of the wavefunction; that would require a higher dimensional Hilbert space than allowed by holographic entropy bounds.

  It  is interesting to consider the consequences of  the idea that  the information deficit  may  actually be observable within our nearly-flat 3D world.  Definition of a classical null trajectory requires localization at both ends, but the precision of transverse localization is limited in the holographic world by the pupil diameter at one end by the diffraction spot size at the other. In particle/wave language, the collapse into a definite classical trajectory occurs with an accompanying uncertainty in angular orientation since the transverse positions with the uncertainty determined by the pupil and diffraction spot are not measured. The sharpest  
 angular information about the 3D world at separation $L$    is encoded so that the  pupil size matches the diffraction spot size, yielding  a characteristic transverse uncertainty $\Delta x_H(L)\simeq \sqrt{Ll_P}$.   An aperture or pupil of diameter $\Delta x_H$ has an angular diffraction limit and uncertainty $\Delta\theta_H\simeq\Delta x_H/L$. In  optics language, one  Planck mass  bends light by an angle $\simeq \Delta\theta_H$ at an impact parameter $\Delta x_H$, bringing it to a focus at $L$, corresponding to one (Planck) cycle of phase difference or time delay across the aperture $\Delta x_H$.  This holographic model suggests that at large separations, Planck-scale quantum gravity effects might be greatly magnified to the scale $\Delta x_H(L)$ by projective holographic optics.

\section{Holographic Uncertainty Principle}
 These considerations motivate a ``holographic uncertainty principle'',  a conjecture about the observable character of quantum indeterminacy of nearly flat spacetime in a particular frame:  {\em  Angular orientations of null trajectories of length  $L$ are uncertain, with standard deviation in each transverse direction:} 
\begin{equation}\label{uncertain}
\Delta \theta>\sqrt{l_P/L}.
 \end{equation}  
 Spatial positions of events with null separation,  at classical positions separated by $\vec L$, are uncertain  in directions transverse to  $\vec L$ with a standard deviation:
 \begin{equation}
 \Delta x_H >\sqrt{Ll_P}.
  \end{equation}
  A more general formulation, in terms of commutation relations between observable position operators at spatially separated events with null separation, is given in  \cite{Hogan:2007ci,Hogan:2007hc}.

This uncertainty can be thought of as a quantum limit to the precision with which parallel rays can be defined.
Angular orientation becomes better defined at larger separation; angles are ill defined at the Planck scale and classical trajectories emerge gradually at macroscopic separation. At   Planck scale separation the world is essentially two-dimensional, defined  in transverse directions at Planck resolution; as larger scales emerge along null trajectories in the third dimension, positions in the transverse directions gradually blur to larger size.

 Formulated this way, the principle makes a Lorentz invariant statement about correlations: all observers on a given branch of the wavefunction can agree on the relationship of (rest frame) $\vec L$ and transverse uncertainty.      Local experiments  retain transverse resolution up to the Planck scale.   However, an experiment conducted in such a way that it is inherently nonlocal--- so that  positions are measured in directions transverse to a large separation, say by a very precise angular measurement, or by  direct transverse position measurements---   reveals  transverse positional uncertainty.
 
  With this uncertainty defining a resolution limit in 3-space, the number of degrees of freedom agrees with holographic bounds. With   Planck scale quanta, a light cone bounded by a sphere of radius $L$  has $\simeq L/l_P$ independent degrees of freedom for each distinguishable direction. Assuming critical sampling,  holographic uncertainty (Eq. \ref{uncertain}) yields $\pi L/l_P$ distinguishable directions in a sphere of radius $L$, yielding $\simeq \pi (L/l_P)^2$ independent degrees of freedom for a light cone bounded by a surface of area $4\pi L^2$, corresponding to the holographic limit. 
   This argument suggests that the holographic uncertainty principle describes conservative departures from classical behavior, in the sense that  observables decohere as little as possible while still respecting holographic bounds.
It is not possible to make more precise relative angular observations because that would require too much information--- more observable eigenvalues could then be measured than allowed by the holographically-bounded size of the Hilbert space.

\section{Direct Measurement of Holographic Noise}
Measurement of holographic uncertainty requires   remote measurement of a very small difference in transverse distance or angle, over large distances or angles.  On the other hand it does not require an experiment at the Planck scale, only a nonlocal comparison that reaches the uncertainty limit, halfway (geometrically) to the Planck scale. It appears that the required sensitivity may be achieved with  appropriately designed interferometers.   A triangle configuration that measures the relative  positions of three proof masses (say)  must be accompanied by holographic  indeterminacy in those positions to agree with holographic entropy  bounds. This can be seen by counting distinguishable spatial pixels since position eigenstates are conjugate to field degrees of freedom. 

The limited amount of information in the system leads to a sampling noise in spatial position observables.   The spectrum of this holographic noise is universal and depends only on the one scale in the system,  $l_P$.  The power spectral density $S_H$ (the mean square dimensionless metric perturbation per frequency interval)  is independent of frequency, and  given at frequency $f\simeq L^{-1}$ simply by\cite{Hogan:2007hc}
\begin{equation}
S_H\simeq  L \Delta \theta^2= l_P,
\end{equation}
with no free parameters.
The rms  amplitude of holographic noise in terms of equivalent metric shear is\cite{Hogan:2007ci}
\begin{equation}\label{noise}
h_{H,rms}=S_H^{1/2}\simeq  \sqrt{l_P/c} = 2.3 \times 10^{-22} /\sqrt{\rm Hz}.
\end{equation} 
These   units, describing the noise due to angular uncertainty, are similar to those used to describe equivalent metric strain noise in gravitational wave detectors, except that the spatial character is a shear rather than a strain.

  One planned experiment that could detect  holographic noise directly at low frequencies is the   Laser Interferometer Space Antenna, { LISA}, which will measure
variations in  distances   between   freely-falling proof masses in three spacecraft forming an approximately equilateral triangle,   $L_{LISA}\simeq5\times 10^6$ km on each side. This configuration allows angular variations to be precisely measured via a Sagnac-type  circuit of all three arms.  
  In its frequency band of maximum sensitivity, from about 3 to 10 millihertz, { LISA}'s measurement accuracy for proof mass position  is limited by system shot noise, with  a reference design noise level \cite{LISA} of $2\times 10^{-9}{\rm cm/\sqrt{Hz}}$. 
The equivalent holographic noise,
$\approx L_{LISA}h_{H,rms}\approx  1\times 10^{-10} {\rm cm/\sqrt{Hz}}$, is significantly below the system noise, but should still be detectable
with  measurements spanning a mission life of several years.
Gravitational-wave noise can be distinguished  from system noise by their different spectral signatures in various linear combinations  of the interferometer phase signals\cite{tinto,Hogan:2001jn,Cornish:2001bb}.  Holographic noise similarly has    a distinctive phase signature 
that allows it to be identified; for example it does  not appear in signals of Michelson-type configurations because they do not measure an angle.

 The  instrument noise of existing ground-based interferometer technology  is   small enough to measure   holographic uncertainty with much higher precision, on a much smaller separation scale, and at much higher frequencies.     Current published data\cite{Abbott:2006zx} shows that LIGO system noise lies below
$h_{rms}= 10^{-22} /\sqrt{\rm Hz}$ 
from about 70 to 400 Hz, well below the predicted level of holographic noise in Eq. (\ref{noise}). Systems with such low noise it should be able to study holographic noise  in detail.  In their current configurations however, ground-based gravitational-wave observatories have  a Michelson layout which  measures the difference between two radial distances from one point, so they have no sensitivity to transverse holographic components of  proof mass position at macroscopic separations.  The spatial structure and frequency spectrum of holographic noise could be studied using a laboratory-scale interferometer with similar laser and proof mass technology to LIGO, but with a design optimized for measuring transverse positions, for example, a closed triangle similar to LISA.  Holographic noise measurements also motivate extension of these studies to  higher frequencies $> >1$ kHz where there are no important gravitational wave sources.   It may be possible with such machines to conduct direct and  detailed measurements of quantum behavior of spacetime.

\acknowledgements
The author is grateful for support from the Alexander von Humboldt Foundation, and for   hospitality from Fermilab, the Kavli Institute for Cosmological Physics, the Enrico Fermi Institute of the University of Chicago, and the Max-Planck-Institut f\"ur Astrophysik.


\begin{thebibliography}{}

\bibitem{Bekenstein:1972tm}
J.~D.~Bekenstein,
Lett.\ Nuovo Cim.\  { 4}, 737 (1972) 

\bibitem{Bardeen:gs}
J.~M.~Bardeen, B.~Carter and S.~W.~Hawking,
Commun.\ Math.\ Phys.\  {\bf 31}, 161 (1973).

\bibitem{Bekenstein:1973ur}
J.~D.~Bekenstein,
Phys.\ Rev.\ D { 7}, 2333 (1973) 

\bibitem{Bekenstein:1974ax}
J.~D.~Bekenstein,
Phys.\ Rev.\ D { 9}, 3292 (1974)

\bibitem{Hawking:1975sw}
S.~W.~Hawking,
Commun.\ Math.\ Phys.\  { 43}, 199 (1975) 







 
\bibitem{'tHooft:1985re}
G.~'t Hooft,
Nucl.\ Phys.\ B { 256}, 727 (1985) 

\bibitem{Strominger:1996sh}
  A.~Strominger and C.~Vafa,
  Phys.\ Lett.\  B {\bf 379}, 99 (1996)

\bibitem{Jacobson:1995ab}
  T.~Jacobson,
  Phys.\ Rev.\ Lett.\  {\bf 75}, 1260 (1995)
  [arXiv:gr-qc/9504004].
  
    
\bibitem{Bombelli:1986rw}
  L.~Bombelli, R.~K.~Koul, J.~H.~Lee and R.~D.~Sorkin,
  Phys.\ Rev.\  D {\bf 34}, 373 (1986).

\bibitem{Srednicki:1993im}
  M.~Srednicki,
  Phys.\ Rev.\ Lett.\  {\bf 71}, 666 (1993)
  [arXiv:hep-th/9303048].
  
\bibitem{Yarom:2004vp}
  A.~Yarom and R.~Brustein,
  Nucl.\ Phys.\  B {\bf 709}, 391 (2005)
  [arXiv:hep-th/0401081].




\bibitem{'tHooft:1993gx}
  G.~'t Hooft,
  arXiv:gr-qc/9310026.


\bibitem{Susskind:1994vu}
  L.~Susskind,
  J.\ Math.\ Phys.\  {\bf 36}, 6377 (1995)
  
\bibitem{'tHooft:1999bw}
G.~'t Hooft,
arXiv:hep-th/0003004.

\bibitem{Bigatti:1999dp}
  D.~Bigatti and L.~Susskind,
  arXiv:hep-th/0002044.
    
\bibitem{Bousso:2002ju}
  R.~Bousso,
  Rev.\ Mod.\ Phys.\  {\bf 74}, 825 (2002)
  
\bibitem{Padmanabhan:2007en}
  T.~Padmanabhan and A.~Paranjape,
  Phys.\ Rev.\  D {\bf 75}, 064004 (2007)
  [arXiv:gr-qc/0701003].
  
\bibitem{Padmanabhan:2007tm}
  T.~Padmanabhan,
  arXiv:0706.1654 [gr-qc].
  
\bibitem{Maldacena:1997re}
  J.~M.~Maldacena,
  Adv.\ Theor.\ Math.\ Phys.\  {\bf 2}, 231 (1998)
  [Int.\ J.\ Theor.\ Phys.\  {\bf 38}, 1113 (1999)]
  
\bibitem{Witten:1998zw}
  E.~Witten,
  Adv.\ Theor.\ Math.\ Phys.\  {\bf 2}, 505 (1998)
  
\bibitem{Aharony:1999ti}
  O.~Aharony, S.~S.~Gubser, J.~M.~Maldacena, H.~Ooguri and Y.~Oz,
  Phys.\ Rept.\  {\bf 323}, 183 (2000)

  
\bibitem{Alishahiha:2005dj}
  M.~Alishahiha, A.~Karch and E.~Silverstein,
  JHEP {\bf 0506}, 028 (2005)
  
\bibitem{Horowitz:2006ct}
  G.~T.~Horowitz and J.~Polchinski,
  arXiv:gr-qc/0602037.


\bibitem{Adelberger:2003zx}
  E.~G.~Adelberger, B.~R.~Heckel and A.~E.~Nelson,
  Ann.\ Rev.\ Nucl.\ Part.\ Sci.\  {\bf 53}, 77 (2003)
  
     
    
\bibitem{Hogan:2007hc}
  C.~J.~Hogan,
  arXiv:0706.1999 [gr-qc].
  
\bibitem{Hogan:2007ci}
  C.~J.~Hogan,
  arXiv:0709.0611 [astro-ph].


  
  \bibitem{LISA}
  {\tt http://lisa.jpl.nasa.gov/facts.html}
  
  \bibitem{tinto}
 M. Tinto, J. W. Armstrong  \& F. B. Estabrook, Phys. Rev. D, 63, 021101 (R)  (2000)

  
\bibitem{Hogan:2001jn}
  C.~J.~Hogan and P.~L.~Bender,
  Phys.\ Rev.\ D {\bf 64}, 062002 (2001)
  
\bibitem{Cornish:2001bb}
  N.~J.~Cornish,
  Phys.\ Rev.\  D {\bf 65}, 022004 (2002)
  





  
  
\bibitem{Abbott:2006zx}
  B.~Abbott {\it et al.}  [LIGO Scientific Collaboration],
  arXiv:astro-ph/0608606.
 
 
 \end{thebibliography}
\end{document}